\documentclass[aps,prd,12pt,showkeys]{revtex4-2}

\usepackage{amsmath}
\usepackage{graphicx}
\usepackage{amsmath, amsthm, amssymb, amsfonts}
\usepackage{dcolumn}
\usepackage{bm}
\usepackage{epsfig}
\usepackage{graphics}
\usepackage{rotating}
\usepackage{xcolor}

\graphicspath{{./figures/}}

\begin{document}


\title{Resolution of the cosmological constant problem by unimodular gravity and signature reversal symmetry}

\author{Recai Erdem}
\email{recaierdem@iyte.edu.tr}
\affiliation{Department of Physics \\
{\.{I}}zmir Institute of Technology\\
G{\"{u}}lbah{\c{c}}e, Urla 35430, {\.{I}}zmir, Turkey}

\date{\today}

\begin{abstract}
The (old) cosmological constant problem consists of two different problems. The first is the huge discrepancy between the value of the cosmological constant deduced from observations and its value expected from cosmological constant-like theoretical contributions (such as vacuum expectation value of Higgs potential). The second problem is why the value of the cosmological constant has its particular (very small) value. It is well-koown that Unimodular gravity solves the first problem while it leaves the second problem unsolved. In this paper I show that the second problem may also be resolved in the context of unimodular gravity by letting our 4-dimensional spacetime be a brane in a $D=2(2n+1)$ dimensional bulk and imposing the signature reversal symmetry. \end{abstract}


\keywords{Cosmological constant problem, unimodular gravity, signature reversal symmetry}

\maketitle

\section{introduction}

 In the standard model of cosmology the accelerated expansion of the universe at the present era is attributed to a positive cosmological constant term in Einstein equations \cite{Weinberg}. The corresponding energy density is measured to be at the order of $\left(10^{-3}~eV\right)^4$. There are many potential theoretical contributions to this energy density that are identified as theoretical contributions to vacuum energy density e.g. zero point energies of particles, vacuum expectation value of Higgs potential, QCD condensate etc.. For example, the expected contribution of Higgs potential to vacuum energy is $\sim\;\left(3\times\,10^{11}~eV\right)^4$, the contribution of QCD vacuum is $\sim\;\left(5\times\,10^{8}~eV\right)^4$, the contribution of zero point energies of quantum fields is expected to be at the order of the fourth powers of masses of the quantum fields \cite{ccp1,zero-point} which, even for electrons results in an energy density of the order of $\sim\;\left(5\times\,10^{5}~eV\right)^4$  (while a different renormalization scheme gives smaller values \cite{Sola,Bull}). In other words, unless one adopts an extreme, unnatural fine-tuned cancellation, the value of the expected theoretical contributions to vacuum energy is extremely larger than its observational value. This is called the (old) cosmological constant problem \cite{Zeldovich,ccp1}. The cosmological constant problem (CCP) persists despite numerous proposals for the solution of the problem \cite{ccp1,ccp2,modified-cc,foam-cc,metric-reversal0} since its identification by Zeldovich in 1967 \cite{Zeldovich}.

 One of the most interesting and well-studied proposals for the solution of CCP is unimodular gravity (UG) \cite{unimodular-gravity,unimodular-approaches,Alvarez-UG-review} . In this setup the contribution of vacuum energy to the cosmological constant vanishes while the cosmological constant survives as an integration constant. In this way CCP is partially solved since the trouble with the huge contributions of the expected theoretical contributions (such as the expectation value of the Higgs potential or QCD vacuum) to CC are evaded. However, CC in this setup is arbitrary, so there is no explanation for its particular phenomenological value. 
 
 Another interesting approach to the solution of CCP is signature reversal symmetry (SRS) \cite{metric-reversal-1,tHooft-Nobbenhuis,metric-reversal-2,Duff,Duff2,metric-reversal-3}. SRS imposes the action functional be invariant under reversal of signature of metric. SRS forbids a cosmological constant term in $D=2(2n+1)$, n = 0,1,2,..... dimensions while the Einstein-Hilbert term (and some matter terms) are allowed. Unlike other symmetries employed for solution of CCP (such as supersymmetry) this symmetry may be taken to be unbroken. However, SRS does not insure vanishing or smallness of CC in dimensions other than $D=2(2n+1)$. One needs somewhat exotic models (as is \cite{metric-reversal-3}) to insure vanishing or smallness of CC in a 3-brane (e.g. in our 4-dimensional spacetime) in the $2(2n+1)$ dimensional bulk. In this way, one may impose SRS to make CC zero and then attribute the accelerated expansion of the universe to other mechanisms such as quintessence \cite{quintessence} or one may let SRS be broken by a small amount to allow a small cosmological constant. In this study I show that UG in conjunction with SRS may resolve CCP in a relatively simple way (without need for a exotic setup).

 In this paper, first a clear description of the (old) CCP is presented in Section II. In Section III the status of CC in UG, and in Section IV SRS are briefly summarized. Then, in Section V a UG approach in conjunction with SRS is adopted to tackle the problem. Finally, we conclude in Section VI.

\section{a clear-cut identification of the old cosmological constant problem}

 The statement of CCP described in literature is somewhat misleading. The potential theoretical contributions to cosmological constant (CC) such as zero point energies of particles, vacuum expectation value of Higgs potential, QCD condensate are not true CC type of contributions. They may be considered as time-dependent CC contributions. The vacuum expectation value of Higgs potential is related to electro-weak phase transition \cite{electroweak}. It is zero before the phase transition and becomes non-zero after the phase transition, so the vacuum expectation value of the Higgs potential is not a true CC contribution. A similar situation holds for QCD condensate. Its contribution is zero before formation of the condensate \cite{phase-transitions,QCD-condensate}. The potential contribution of zero-point energies (that are proportional to the particle masses) vanish before the electro-weak phase transition since the Higgs field expectation value (that induces particle masses) vanishes before the phase transition. Cosmological constant (in general relativity) by definition has the same constant value everywhere in the universe for all times while the potential contributions to CC mentioned above are time-dependent. In other words, these potential contributions, in a strict sense, are not true CC type contributions. This point will be important in the following analysis.

\section{unimodular gravity and the old cosmological constant problem}

UG solves the old cosmological constant problem in the sense that the expected theoretical contributions to CC (such as zero point energies of particles, vacuum expectation value of Higgs potential, QCD condensate) at the classical field level do not weigh in UG while the true CC (that arises as an integration constant in UG) survives. This point will be emphasized and summarized below. Note that the capacity of UG to resolve CCP after quantum corrections is a debated issue, and probably depends on the unimodular approach to UG \cite{unimodular-approaches} in the case of quantization of matter sector or to quantization procedure in the case quantum corrections to gravitational sector \cite{WTDiff-quantum}). The analysis below will be in the context of classical field theory (while the approach followed below is considered as a safe approach to UG in the context of quantum corrections by matter fields in \cite{unimodular-approaches}, and the quantized version of the gravitational sector in UG is shown to be equivalent to general relativity in \cite{WTDiff-quantum}).

 We consider the following action \cite{unimodular-gravity}
\begin{equation}
S\,=\,S_{g}\,+\,S_m \label{eq1}
\end{equation}
Here
\begin{equation}
~~~S_g\,=\,\frac{1}{2\kappa^2}
\int\,d^Dx\;\omega\,\left[R(\bf{\tilde{g}})\rm\,-2\Lambda\right]~,~~~ S_m\,=\,
\int\,d^Dx\;\omega\,{\cal L}_m \label{eq2}
\end{equation}
where $\Lambda$ is a constant, $\bf{\tilde{g}}$ denotes the metric tensor. Determinant of $\bf{\tilde{g}}$ is denoted by $det\left(\bf{\tilde{g}}\rm\right)$ with $\omega=\sqrt{|\left(\bf{\tilde{g}}\rm\right)|}$ being assumed to be fixed i.e. it does not vary under functional variations, so $\omega$ does not contribute to the classical field equations. The corresponding diffeomorphisms and Weyl scalings of the metric $\bf{\tilde{g}}$ are restricted to those that keep $\omega$ fixed, namely, to transverse diffoemorhisms (TDiff) i.e. to the diffeomorphisms with $\tilde{\nabla}_A\xi^A=0$ where $\tilde{\nabla}_A$,
$A = 0,1,....,D-1$, denotes the covariant derivative with respect $\bf{\tilde{g}}$ and $\xi^A$ is a vector field that keeps $\omega$ fixed. 
$S_m$ is the action for matter fields that relates to energy momentum tensor $T_{AB}$ by
\begin{equation}
T_{AB}(\bf{\tilde{g}}\rm)\,=\,\frac{-2}{\sqrt{|det\left(\tilde{g}\right)}|}\frac{\delta S_m}{\delta\tilde{g}^{AB}}  \label{eq3}
\end{equation}

The condition of $\omega^2=|det\left(\tilde{g}\right)|$ being fixed (under functional variations), namely, $\delta\,\left(det\tilde{g}\right)=0$ (i.e. $\tilde{g}_{AB}\,\delta \tilde{g}^{AB}=0$) implies
\begin{equation}
\delta \tilde{g}^{AB}\,=\,\delta g^{AB} - \frac{1}{D}\,g^{AB}g_{CF}\;\delta g^{CF}          \label{eq4}
\end{equation}
and
\begin{equation}
\tilde{g}_{AB}\,=\,g_{AB}\,\left(\frac{\omega^2}{|g|}\right)^\frac{1}{D}.        \label{eq5}
\end{equation}
where $g_{AB}$ is a generic metric tensor, and $|g|=|det\left(\bf{g}\rm\right)|$. (Use of $\tilde{g}_{AB}$ rather than $g_{AB}$ corresponds to extension TDiff to WTDiff with inclusion of Weyl transformations \cite{WTDiff,unimodular-gravity,WTDiff-quantum}, and provides a simple way of imposing unimodularity condition).
The field equations corresponding to (\ref{eq2}) may be easily obtained from variation of $S$ by making use of (\ref{eq4}) and they turn out to be
\begin{equation}
R_{AB}(\tilde{\bf{g}}\rm)\,-\,\frac{1}{D}\tilde{g}_{AB}\,R(\tilde{\bf{g}}\rm)\,=\,\kappa^2\,\left(T_{AB}(\tilde{\bf{g}}\rm)
\,-\,\frac{1}{D}\tilde{g}_{AB}\,T(\tilde{\bf{g}}\rm)\right).  \label{eq6}
\end{equation}
Eq.(\ref{eq6}) is the gravitational field equation for unimodular gravity.

 By using the Bianchi identity $\tilde{\nabla}_A\left[R^{AB}(\tilde{\bf{g}}\rm)\,-\,\frac{1}{2}\tilde{g}^{AB}\,R(\tilde{\bf{g}}\rm)\,\right]=0$ and conservation of energy-momentum tensor $\tilde{\nabla}_A\,T^{AB}(\tilde{\bf{g}}\rm)\,=\,0$ in (\ref{eq6}) one may write \cite{Fabris}
\begin{equation}
\frac{D-2}{2D}R(\tilde{\bf{g}}\rm)\,+\,\frac{1}{D}\kappa^2\,T(\tilde{\bf{g}}\rm)
\,=\,-\tilde{\Lambda}  \label{eq6a}
\end{equation}
where $\tilde{\Lambda}$ is an arbitrary constant. After using (\ref{eq6a}) in (\ref{eq6}), one may derive the Einstein equation of general relativity, namely, \begin{equation}
R_{AB}(\tilde{\bf{g}}\rm)\,-\,\frac{1}{2}\,\tilde{g}_{AB}\,R(\tilde{\bf{g}}\rm)\,+\,\tilde{g}_{AB}\tilde{\Lambda}\,=\,\kappa^2\,T_{AB}(\tilde{\bf{g}}\rm)
 \label{eq6aa}
\end{equation}
One notices from  (\ref{eq6aa}) that $\tilde{\Lambda}$ is the Einstein's cosmological constant.

Note that the action (\ref{eq1}) with (\ref{eq2}) is the same as the standard UG action except the additional redundant $-2\Lambda$ term. This term is not present in the original unimodular gravity action because it does not contribute to the gravitational field equations of UG (\ref{eq6}). (However, this term will play an important role in the modified scheme that we introduce in Appendix A). In fact, we may let $\Lambda$ to have a more general form than being a constant. In general, $\Lambda$ may be a scalar function of coordinates that does not depend on metric tensor. Even in that case $\Lambda$ does not contribute to field equations of unimodular gravity. For example, we may let $2\Lambda$ be the potential term of a scalar field (provided that ${\cal L}_m$ denotes the Lagrangian for the matter fields except the potential term). In literature such terms are called unimodular ambiguity \cite{Tiwari,Finkelstein}. The cosmological constant like contributions such as zero point energies of particles, vacuum expectation value of Higgs potential, QCD condensate are unimodular ambiguity type contributions, so they do not weigh in UG. The true cosmological constant is the $\tilde{\Lambda}$ in (\ref{eq6aa}). In other words, the cosmological constant-like terms such as zero point energies of particles, vacuum expectation value of Higgs potential, QCD condensate do not contribute to the gravitational field equation of UG i.e. they do not weigh while the true cosmological constant $\tilde{\Lambda}$ survives in UG as an arbitrary integration constant.

\section{Signature reversal symmetry (SRS)}

Signature reversal flips the sign of metric \cite{metric-reversal-1,tHooft-Nobbenhuis,metric-reversal-2,Duff} i.e.
\begin{equation}
ds^2\,=\,g_{AB}dx^Adx^B~\rightarrow~-ds^2~~~~A,B=0,1,...D-1 \label{srs1}
\end{equation}
which may be induced by \cite{metric-reversal-1,tHooft-Nobbenhuis}
\begin{equation}
x^A~\rightarrow~i\,x^A~~ \mbox{and}~~ g_{AB}~\rightarrow~\,g_{AB}  \label{srs2a}
\end{equation}
 or by \cite{metric-reversal-2,Duff}
\begin{equation}
x^A~\rightarrow~\,x^A~~ \mbox{and}~~ g_{AB}~\rightarrow~-\,g_{AB}  \label{srs2}
\end{equation}

If (\ref{srs2a}) or (\ref{srs2}) is imposed at the level of Einstein equations, then, after requiring all terms transform in the same way under the transformation, a cosmological constant is forbidden in any dimension. (\ref{srs2a}) or (\ref{srs2}) may be also imposed to leave the gravitational action invariant (so that it becomes also relevant at quantum level).
One observes that, in $D=2(2n+1)$, $n=0,1,2...$ dimensional spaces, under either realization of the symmetry we have
\begin{equation}
R\,\rightarrow\,-\,R~,~~\sqrt{|det(g)|}\;d^Dx\,\rightarrow\,-\,\sqrt{|det(g)|}\;d^Dx~~\mbox{where}~D=2(2n+1). \label{srs3}
\end{equation}
(Here, by $|det(g)|$ we imply that $|det(g)|$ is always taken to be positive (rather than implying a true absolute value) i.e. $|det(g)|=det(g)$ when ${\cal S}\,-\,{\cal T}$ = even while $|det(g)|=-\,det(g)$ when ${\cal S}\,-\,{\cal T}$ = odd where ${\cal S}$ and ${\cal T}$ denote the number of space-like and time-like coordinates.)
while $S_m$ in (\ref{eq1}) (so $T$ in (\ref{eq6a})) may have both and even parts in general i.e. $T\,=\,T^(e)\,+\,T^{(o)}$ with 
\begin{equation}~~
	T^{(e)}\,\rightarrow\,T^{(e)}~,~~
	T^{(o)}\,\rightarrow\,-T^{(0)} ~~~\mbox{as}~~~g_{AB}~\rightarrow~-\,g_{AB}. \label{srs3x}
\end{equation}

Eqs. (\ref{srs3}) and (\ref{srs3x}) imply that the Einstein-Hilbert action and $T^{(o)}$ are allowed while CC and $T^{(e)}$  (e.g. potential terms that do not depend on metric tensor) are not allowed in $D=2(2n+1)$ dimensions by SRS. (Note that there are realization of SRS such as (\ref{srs2}) supplemented by reversal of string coupling constants \cite{Duff2} or by additional discrete symmetries \cite{metric-reversal-3} but the main implications of all these realizations of SRS are essentially the same).

In \cite{tHooft-Nobbenhuis} it was shown that imposing (\ref{srs2a}) on Einstein equations (in any dimension) implies that it is a symmetry of gravitational sector with a vanishing CC i.e. it is a symmetry of vacuum \cite{tHooft-Nobbenhuis}. On the other hand, (\ref{srs2}) imposed at the level of action functional in $2(2n+1)$ dimensions seems to have a richer spectrum \cite{metric-reversal-3} (while in a somewhat exotic setup). In either case SRS is a symmetry of vacuum.  In the light of the studies on emergence of signature \cite{signature-change} and emergence of universe from nothing \cite{emergence1,emergence2,emergence3}, one may speculate that SRS is a result of a universe where initially (in a pre-geometry era) all signatures were equally possible and there were random signature transitions between different signatures (including signature reversals), and then the universe settled to one of the signatures e.g. by spontaneous symmetry breaking \cite{speculation}.

A comment is in order here; the infinitesimal volume element $\sqrt{|det(g)|}\;d^Dx$ in (\ref{srs3}) may be replaced by a non-metric volume element, say $\Phi\,d^4x$, in a modified gravity context \cite{Guendelman1,Guendelman2,Guendelman3}. However, one should be cautious when the results of \cite{Guendelman1,Guendelman2,Guendelman3} are compared with the present study and with (\cite{metric-reversal-1},\cite{metric-reversal-2},\cite{metric-reversal-3}). The transformations in \cite{Guendelman1,Guendelman2,Guendelman3} are considered as diffeomorphisms (in a given spacetime manifold) while in (\cite{metric-reversal-1},\cite{metric-reversal-2},\cite{metric-reversal-3}) they are imposed as symmetries between two universes with opposite signatures. However, one may also impose similar symmetries to the models with non-metric volume elements.  In that case, depending on how $\Phi$ transforms, one may get the same or different conclusions after imposing invariance of the action. For example if $\Phi\,d^4x$ and $R$ transform in the same way as (\ref{srs3}), then the conclusions remain the same while otherwise they may differ. There is a common element in both sets of studies. The transformations in \cite{Guendelman1,Guendelman2,Guendelman3} are considered as local transformations that do not include the boundaries of the spacetime manifold \cite{Guendelman2,Guendelman3}. The transformation in (\ref{srs1}) may also be considered to be local in the sense that it is imposed at the beginning of the universe in pre-geometric era. Moreover, an elegant formulation requires introduction of the concept of signed transformations that, in turn, requires extension of the spacetime manifold to a complexified one \cite{Guendelman1,Guendelman2}. In fact, the same extension is implied by the (\ref{srs2a}) realization of SRS. It seems that, in this way, the $ds^2$ and $-ds^2$ may be included in a complex manifold in a consistent way. All these points need detailed studies in future. In the next section I will simply assume that SRS is a symmetry of nature and try to see its impact on CCP in a UG framework.

\section{unimodular gravity in conjunction with signature reversal symmetry resolves the cosmological constant problem}

In the section before the preceding section ,we have seen that unimodular ambiguity 
(so, vacuum energy) does not weigh in UG, however, CC remains a free parameter. On the other hand the results of the preceding section implies that, if we let our 4-dimensional spacetime be a 3-brane in a $2(2n+1)$-dimensional bulk and impose SRS, then the bulk CC vanishes while a non-vanishing 4-dimensional CC in the 3-brane is allowed \cite{Duff}. Therefore, to solve CCP, a mechanism that makes the 4-dimensional CC vanish (so that dark energy may attributed to an alternative mechanism such as quintessence \cite{quintessence}) or that induces a CC at correct order of magnitude should be found. In \cite{metric-reversal-3} this has been tried to be accomplished through a complicated way. Below we will see that CPP may be resolved in a simple way by employing UG in conjunction with SRS. In the following subsection, we will consider the case where SRS in the gravitational sector and transverse diffeomorphism are exact. Then, in the second subsection we  will discuss the situation where SRS is broken by a modified gravity contribution so that a CC emerges at the right order of magnitude (i.e. at the order of the total energy density of the universe). This discussion will also illuminate the cases of broken and unbroken SRS in a more concrete way. Another approach may be breaking TDiff by a small amount. This last approach is discussed in Appendix A.

\subsection{The case of unbroken SRS and TDiff}
If we impose SRS and assume UG in a $2(2n+1)$-dimensional bulk, then the CC of a brane in the bulk also vanishes since a brane CC is simply a constant times a delta function that confine the CC to the brane, hence it is a unimodular ambiguity that does not give a contribution to the gravitational field equations of UG.  

To see the situation more clearly let us next consider the integration constant $\tilde{\Lambda}$ in this framework. Under an SRS transformation Eq.(\ref{eq6a}) goes to
\begin{equation}
	-\frac{D-2}{2D}R(\tilde{\bf{g}}\rm)\,+\,\frac{1}{D}\kappa^2\,\left[T^{(e)}(\tilde{\bf{g}}\rm)\,-\,T^{(o)}(\tilde{\bf{g}}\rm) \right]
	\,=\,-\tilde{\Lambda}  \label{eq6ay}
\end{equation}
Note that the right hand side does not change sign since it is a constant. Eq.(\ref{eq6a}) and Eq.(\ref{eq6ay}) together imply that
\begin{equation}
	\frac{D-2}{2D}R(\tilde{\bf{g}}\rm)\,+\,\frac{1}{D}\kappa^2\,\,T^{(o)}(\tilde{\bf{g}})
\,=\,0
~~~~\mbox{\it{and}}~~~~~ \frac{1}{D}\kappa^2\,T^{(e)}(\tilde{\bf{g}}\rm)\,=\,\tilde{\Lambda}\,\neq\,0.
	\label{x1}
\end{equation} 
The first case above with $\tilde{\Lambda}=0$ corresponds to the case of an exact SRS. The seconds case with $\tilde{\Lambda}\,\neq\,0$ corresponds to breaking of SRS, and, in fact, is a trivial case of a cosmological constant type of term being the only contribution to the energy-momentum tensor. It is evident that this case does not correspond to the universe at present. In the first case the dark energy may be identified by a quintessence like term. We will see below that the second option becomes non-trivial when SRS is broken by inclusion of a modified gravity term.      

\subsection{The case of breaking SRS by a small amount}
We assume unimodularity in a  $2(2n+1)$-dimensional bulk and on our 4-dimensional brane, and break SRS by a Starobinsky type term, namely, we keep the $S_m$ term in (\ref{eq1}) the same while the $S_g$ term is modified as follows 
\begin{equation}
	~~~S_g\,=\,\frac{1}{2\kappa^2}
	\int\,d^Dx\;\omega\,\left[R(\bf{\tilde{g}})\rm\,+\,\alpha\,R^2(\bf{\tilde{g}})\rm)-2\Lambda\right],  \label{eq2z}
\end{equation}
where $\alpha$ is constant. The metric tensor may be compactified along the extra-dimensions i.e.
\begin{equation}
\bf{\tilde{g}}\rm_{AB}\,=\,\bf{\tilde{g}}\rm_{AB}^{(0)}\,+\,\sum_a\,\bf{\tilde{g}}\rm_{AB}^{(a)}  \label{eq2za}
\end{equation}
where $\bf{\tilde{g}}\rm_{AB}^{(0)}$ stands for the zero mode and $\bf{\tilde{g}}\rm_{AB}^{(a)}$ for higher modes.
 We let $\bf{\tilde{g}}\rm_{\mu\,a}=\bf{\tilde{g}}\rm_{a\,\mu}=0$ with $\mu=0,1,2,3$, $a=4,...D-1$. Further, we let $\bf{\tilde{g}}\rm_{\mu\nu}$ may depend on extra-dimensions only through conformal factors which is an assumption that may be justified by physical arguments. Then, we have
\begin{equation}
	R(\bf{\tilde{g}})\,=\,R^{(4)}\,+\,\Delta_R~,~~~det\left(\bf{\tilde{g}}\rm_{AB}\right)\,=\,\beta\,det\left(\bf{\tilde{g}}\rm_{\mu\nu}^{(0)}\right)\,
	det\left(\bf{\tilde{g}}\rm_{ab}\right),
	\end{equation} 
	where $R^{(4)}$ depends only on $\bf{\tilde{g}}\rm_{\mu\nu}^{(0)}$, and $\Delta_R$ denotes the rest of the terms in $R(\bf{\tilde{g}})$, $\beta$ depends only on extra dimensions and is due to a possible extra-dimensional conformal factor of $\bf{\tilde{g}}\rm_{\mu\nu}$. Hence, (\ref{eq2z}) may be expressed as
\begin{eqnarray}
	~~~S_g&=&\frac{1}{2\kappa^2}
	\int\,d^Dx\;\omega\,\left[R^{(4)}\,+\,\alpha\,\left(R^{(4)}\right)^2-2\Lambda\right]\,+\,	\int\,d^Dx\;\omega\,\Delta  \nonumber \\
	&&=\,	\frac{1}{2\kappa_4^2}\,\int\,d^4x\;\omega_4\,\,\left[R^{(4)}\,+\,\alpha\,\left(R^{(4)}\right)^2-2\Lambda\right]\,+\,
	\int\,d^Dx\;\omega\,\Delta 
	\label{eq2zz}
\end{eqnarray}
where $\Delta$ denotes the rest of the terms (that may also depend on extra dimensions), and 	$\frac{1}{2\kappa_4^2}\,=\,\frac{1}{2\kappa^2}\int\,d^{D-4}x\;\beta\omega_d$, $\omega_4^2=|det\left(\tilde{g}_{\mu\nu}^{(0)}\right)|$, $\omega_d^2=|det\left(\tilde{g}_{ab}\right)|$.

We assume $\omega$ and $\omega_4$ being fixed under variation i.e. we assume unimodularity condition for the $D$ dimensional bulk and for our 4-dimensional spacetime. Hence, the $D$-dimensional gravitational field equations corresponding to (\ref{eq2z}) and the 4-dimensional gravitational field equations corresponding to (\ref{eq2zz}) are \cite{Eichorn,MOG-UMG}
\begin{eqnarray}
	&&(1+2\alpha\,R)\left(R_{AB}(\tilde{\bf{g}}\rm)\,-\,\frac{1}{D}\tilde{g}_{AB}\,R(\tilde{\bf{g}}\rm)\right)
	\,+\,2\alpha\left(\frac{1}{D}\tilde{g}_{AB}\tilde{\Box}\,-\,\tilde{\nabla}_A\tilde{\nabla}_B\right)\,R 
	\,+\,2\alpha\frac{1-D}{D}\,\tilde{g}_{AB}\,\tilde{\Box}\,R(\tilde{\bf{g}}) \nonumber \\
	&&=\,\kappa^2\,\left(T_{AB}(\tilde{\bf{g}}\rm)
	\,-\,\frac{1}{D}\tilde{g}_{AB}\,T(\tilde{\bf{g}}\rm)\right)
	\label{eq6x1}
\end{eqnarray}
\begin{eqnarray}
	&&(1+2\alpha\,R^{(4)})\left(R_{\mu\nu}^{(4)}(\tilde{\bf{g}}\rm)\,-\,\frac{1}{D}\tilde{g}^{(0)}_{\mu\nu}\,R^{(4)}(\tilde{\bf{g}}\rm)\right)
\,+\,2\alpha\left(\frac{1}{4}\tilde{g}^{(0)}_{\mu\nu}\tilde{\Box}^{(4)}\,-\,\tilde{\nabla}_\mu\tilde{\nabla}_\nu\right)\,R^{(4)}
\,-\,\frac{3}{2}\alpha\,g_{\mu\nu}\,\Box^{(4)}\,R^{(4)}(\tilde{\bf{g}})\,\nonumber \\
&&=\,\kappa_4^2\,\left(T_{\mu\nu}^{(4)}(\tilde{\bf{g}}\rm)
\,-\,\frac{1}{D}\tilde{g}^{(0)}_{\mu\nu}\,T^{(4)}(\tilde{\bf{g}}\rm)\right)
	\label{eq6x2}
\end{eqnarray}
where $R^{(4)}$, $R_{\mu\nu}^{(4)}$, $\tilde{g}^{(0)}_{\mu\nu}$, $\tilde{\Box}^{(4)}$, $\tilde{\nabla}_\mu$, $T^{(4)}$, $T_{\mu\nu}^{(4)}$ are expressed in terms of zero modes, so do not depend on extra dimensions, and 
\begin{equation}
T_{\mu\nu}^{(4)}(\bf{\tilde{g}}\rm)\,=\,\frac{-2}{\sqrt{\left|det\left(\tilde{g}^{(0)}_{\mu\nu}\right)\right|}}\frac{\delta\left[S_m\,+\,\int\,d^Dx\;\omega\,\Delta \right]}{\delta\,\tilde{g}^{(0)\mu\nu}}  \label{eq3x}
\end{equation}

After taking the divergence of both sides of the equations and using Bianchi identities and conservation of energy-momentum tensor and using $\left(\Box\nabla_A-\nabla_A\Box\right)F=R_{AB}\nabla^BF$ \cite{Koivisto,MOG-UMG} (that may be obtained from $[\nabla_A,\nabla_B]V_C=R^G_{ABC}\,V_G$) for any scalar $F$; Eqs. (\ref{eq6x1}) and  (\ref{eq6x2}) result in 
\begin{equation}
	\frac{D-2}{2D}R(\tilde{\bf{g}}\rm)\,-\,\frac{2\alpha}{D}\left(R(\tilde{\bf{g}}\rm)\right)^2\,+\,\frac{1}{D}\kappa^2\,T(\tilde{\bf{g}}\rm)
		\,+\,2\alpha\frac{1-D}{D}\,\tilde{\Box}\,R(\tilde{\bf{g}})\,=\,-\tilde{\Lambda}  \label{eq6ax1}
\end{equation}
\begin{equation}
	\frac{1}{4}\,R^{(4)}(\tilde{\bf{g}}\rm)\,-\,	\frac{\alpha}{2}\left(R^{(4)}(\tilde{\bf{g}}\rm)\right)^2\,+\,\frac{1}{4}\kappa_4^2\,T^{(4)}(\tilde{\bf{g}}\rm)
	\,-\,\frac{3}{2}\alpha\,\Box^{(4)}\,R^{(4)}(\tilde{\bf{g}})
	\,=\,-\tilde{\Lambda}_4  \label{eq6ax2}
\end{equation}

One may write $T_{AB}\,=\,T^{(o)}_{AB}+T^{(e)}_{AB}$, $T^{(4)}_{\mu\nu}\,=\,T^{(4)\,(o)}_{\mu\nu}+T^{(4)\,(e)}_{\mu\nu}$, where the upper scripts $(o)$ and $(e)$ denote the part of the quantity that change or do not change sign under SRS transformations given by (\ref{srs2}). We apply (\ref{srs2}) to (\ref{eq6ax1})  and (\ref{eq6ax2}), and add and subtract the resultant equations to the original equations. Then,  (\ref{eq6ax1})  and (\ref{eq6ax2}) result in 
\begin{equation}
	-\frac{2\alpha}{D}\left(R(\tilde{\bf{g}}\rm)\right)^2\,+\,\frac{1}{D}\kappa^2\,T^{(e)}(\tilde{\bf{g}}\rm)	\,+\,2\alpha\frac{1-D}{D}\,\tilde{\Box}\,R(\tilde{\bf{g}})
	\,=\,-\tilde{\Lambda}  \label{eq6axx1a}
\end{equation}
\begin{equation}
	\frac{(D-2)}{2D}\,R(\tilde{\bf{g}}\rm)\,+\,\frac{1}{D}\kappa^2\,T^{(o)}(\tilde{\bf{g}}\rm)	
	\,=\,0  \label{eqx6axx1b}
\end{equation}
\begin{equation}
	-\frac{\alpha}{2}\left(R^{(4)}(\tilde{\bf{g}}\rm)\right)^2\,+\,\frac{1}{4}\kappa_4^2\,T^{(4)\,(e)}(\tilde{\bf{g}}\rm)
	\,-\,\frac{3}{2}\alpha\,\Box^{(4)}\,R^{(4)}(\tilde{\bf{g}})
\,=\,-\tilde{\Lambda}_4  \label{eq6axx2a}
\end{equation}
\begin{equation}
R^{(4)}(\tilde{\bf{g}}\rm)\,+\,\kappa_4^2\,T^{(4)\,(o)}(\tilde{\bf{g}}\rm)\,=\,0  \label{eqx6axx2b}
\end{equation}

Eq.(\ref{eqx6axx1b}) is the same as the first equation of (\ref{x1}). This is expected since the quadratic gravity term introduced here is even under SRS, so contributes only to the second equation of (\ref{x1}). Although the additional terms in (\ref{eq6axx1a}) seem as a straightforward extension of the second equation of (\ref{x1}), in fact they are more than that. Now, the $T^{(e)}$ term is not necessarily of the form of a cosmological constant term. It may be a general SRS even term in a Lagrangian e.g. a potential term for a scalar and/or a kinetic term for a gauge field. The equations
(\ref{eq6axx2a}) and (\ref{eqx6axx2b}) have similar implications for our 4-dimensional spacetime, and they are, in fact, are the main equations that we are interested in. Eq. (\ref{eq6axx2a}) is especially important. It tells us that a nontrivial cosmological constant $\tilde{\Lambda}_4$ is possible in the presence of a quadratic gravity term even when $T^{(4)\,(e)}$ is zero. 
 
\section{conclusion}

The incentive for the present study follows from three observations. The first observation is that unimodular gravity is successful for the resolution of the most impressing part of the old cosmological constant problem, namely, unimodular ambiguity (so its particular form vacuum energy) does not gravitate (while cosmological constant survives as an arbitrary integration constant) in unimodular gravity. The second observation motivating the present study is that signature reversal symmetry is an exact symmetry of Einstein-Hilbert action that forbids a bulk cosmological constant in $D=2(2n+1)$ dimensions. The third observation is that brane cosmological constants in a bulk have the form of unimodular ambiguity. The consequence of these three observations turned to be the following conclusion: cosmological constant problem may be resolved if we assume that we live on a 3-brane in a $D=2(2n+1)$ dimensional bulk in a unimodular gravity framework. Therefore, first, the essential points of unimodular gravity and signature reversal symmetries in connection with the present study are briefly summarized. Then, the main topic of this study, namely, vanishing of cosmological constant and vacuum energy in a universe that corresponds to a 4-dimensional brane in a $D=2(2n+1)$ dimensional bulk with unimodular gravity are considered. It is found that the cosmological constant and vacuum energy vanish in such a setting if SRS is an exact symmetry. The accelerated expansion of the universe in this setup may be attributed to a mechanism other than cosmological constant or to a small violation of transverse diffeomorphims as discussed in Appendix A. Moreover, violation of signature reversal symmetry by an additional modified gravity term has also been studied in the simplest case of adding a $R^2$ term to Einstein-Hilbert action. In this case, a non-vanishing cosmological constant emerges in a non-trivial way. Note that employing other types of modified gravity terms may also be interesting in this context. In particular, theories with non-Riemannian volume elements \cite{Guendelman1,Guendelman2,Guendelman3} may be interesting in the view that Henneaux-Teitelboim gravity \cite{Henneaux-Teitelboim} is a unimodular gravity framework with non-metric volume element $\Phi\,d^4x$ (instead of $\sqrt{|det(\bf{g})}|d^4x$). (In this theory, unimodularity condition is imposed in a general covariant framework by gauge fixing $\Phi$ to a volume density $\omega$). In the light of these observations it may be interesting in future to consider extensions of the present study along the above mentioned lines.

\appendix

\section{A small CC through SRS in conjunction with an extremely small TDiff violation}

Let us replace the unimodular gravity condition $\frac{\delta\omega}{\delta\tilde{g}_{\mu\nu}}=0$ i.e. $\tilde{g}_{\mu\nu}\,\delta\tilde{g}^{\mu\nu}=\,0$ by
\begin{equation}
\,\tilde{g}_{\mu\nu}\,\delta\tilde{g}^{\mu\nu}\,=\,\epsilon_1\,g_{\mu\nu}\,\delta\,g^{\mu\nu}.  \label{eq7}
\end{equation}
where $\epsilon_1$ is a very small constant, and its smallness may be justified by violation of an extremely small TDiff  violation. One may speculate that the non-vanishing of the right-hand side of (\ref{eq7}) may be considered to be due to some unknown quantum gravitational effects.
The easiest way to impose (\ref{eq7}) in field equations is to modify (\ref{eq4}) as
\begin{equation}
\delta \tilde{g}^{\mu\nu}\,=\,\delta g^{\mu\nu} - \frac{1}{4}\,g^{\mu\nu} g_{\alpha\beta}\;\delta g^{\alpha\beta} -
\frac{\epsilon_1}{2}g^{\mu\nu} \left(\frac{\omega^2}{|g|}\right)^{-\frac{1}{4}} g_{\alpha\beta}\;\delta g^{\alpha\beta}       \label{eq4a}
\end{equation}
while (\ref{eq5}) remains the same. Thus, (\ref{eq7}) implies
\begin{equation}
\delta \omega\,=\,-\frac{1}{2} \omega \tilde{g}_{\mu\nu}\,\delta\tilde{g}^{\mu\nu}\,=\,-\frac{\epsilon_1}{2}\omega\, g_{\mu\nu}\,\delta\,g^{\mu\nu}.
\label{eq4a}
\end{equation}
Hence, (\ref{eq6}) is replaced by
\begin{equation}
R_{\mu\nu}(\bf{g}\rm)\,-\,\frac{1}{4}\left(1+4\epsilon_1\right)\,g_{\mu\nu}\rm\,R(\bf{g})\rm\,-\frac{1}{2}\epsilon_1\,g_{\mu\nu}\Lambda\,=\,\kappa^2\,
\left(T_{\mu\nu}(\bf{g}\rm)\,-\,\frac{1}{4}\left(1+2\epsilon_1\right)g_{\mu\nu}\,T(\bf{g}\rm)\,\right)
\label{eq8}
\end{equation}
By Bianchi identity and conservation of energy-momentum tensor, (\ref{eq8}) results in
\begin{equation}
\left[\left(1-4\epsilon_1\right)\,R(\bf{g}\rm)\,-\,2\epsilon_1\Lambda+\,\kappa^2\,\left(1+2\epsilon_1\right)\,T(\bf{g}\rm)\right]_{,\mu}\rm\,=\,0  \label{eq9}
\end{equation}
which implies
\begin{equation}
\left(1-4\epsilon_1\right)\,R(\bf{g}\rm)\,-\,2\epsilon_1\Lambda+\,\kappa^2\,\left(1+2\epsilon_1\right)\,T(\bf{g}\rm)\,=\,-4\tilde{\Lambda}  \label{eq10}
\end{equation}
where the integration constant $\tilde{\Lambda}$ is identified as the cosmological constant. Eq.(\ref{eq8}), after the use of (\ref{eq10}), becomes
\begin{equation}
R_{\mu\nu}(\bf{g}\rm)\,-\,\frac{1}{2}g_{\mu\nu}\,R(\bf{g}\rm)\,-\,g_{\mu\nu}\tilde{\Lambda}\,=\,\kappa^2\,
T_{\mu\nu}(\bf{g}\rm)
\label{eq11}
\end{equation}
which is the Einstein equation with cosmological constant $\tilde{\Lambda}$ as expected.

Finally, I will show that $\tilde{\Lambda}$  is very small provided that $\epsilon_1$ is a very small. Applying SRS transformation to both sides of (\ref{eq9}) result in an equation that may be obtained from (\ref{eq9}) by flipping the signs of the $R(\bf{g})$ and $T(\bf{g}$) terms  while the $\Lambda$ and $\tilde{\Lambda}$ terms remain the same. After taking the sum of the resultant equation and (\ref{eq9}) we get
\begin{equation}
\tilde{\Lambda}\,=\,\frac{1}{2}\epsilon_1\Lambda \label{eq12}
\end{equation}
i.e. $\tilde{\Lambda}$ can be made sufficiently small by taking $\epsilon_1$ sufficiently small provided that $\Lambda$ is finite.



\end{document}